\documentclass[10pt,letterpaper]{article}
\usepackage{opex3}

\graphicspath{{figures/}}

\newcommand{\e}{\varepsilon}
\newcommand{\m}{\mu}
\newcommand{\w}{\omega}
\newcommand{\vc}[1]{\mathbf{#1}}
\newcommand{\uv}[1]{\vc{\hat{#1}}}

\usepackage{amsmath}
\DeclareMathOperator{\sgn}{sgn}

\begin{document}

\title{All-angle collimation of incident light in $\vc{\m}$-near-zero
       metamaterials}

\author{Vladimir~Yu.~Fedorov$^*$ and Takashi~Nakajima$^1$}

\address{Institute of Advanced Energy, Kyoto University, Gokasho, Uji, Kyoto
         611-0011, Japan \\
        $^1$nakajima@iae.kyoto-u.ac.jp}

\email{$^*$v.y.fedorov@gmail.com}

\begin{abstract}
    We use the theory of inhomogeneous waves to study the transmission of light
    in $\m$-near-zero metamaterials. 
    We find the effect of all-angle collimation of incident light, which means
    that the vector of energy flow in a wave transmitted to a $\m$-near-zero
    metamaterial is perpendicular to the interface for any incident angles if an
    incident wave is s-polarized.
    This effect is similar to the all-angle collimation of incident light
    recently found through a different theoretical framework in $\e$-near-zero
    metamaterials for a p-polarized incident wave [S.~Feng, \prl {\bf 108,}
    193904 (2012)].
    To provide a specific example, we consider the transmission of light in a
    negative-index metamaterial in the spectral region with a permeability
    resonance, and show that all-angle collimation indeed takes place at the
    wavelength for which the real part of permeability is vanishingly small.
\end{abstract}


%********************************************************************************
\section{Introduction}
%********************************************************************************

Refraction of light is the fundamental optical phenomenon.
Significant progress in fabrication of nanoscale structures led to creation of
optical metamaterials which allow us to manipulate the way light refracts.
For example, at the interface of negative-index metamaterials, the angle of
refraction turns out to be negative \cite{2001_Shelby,2007_Lezec}.
In~\cite{2011_Yu}, an array of optically thin resonators with subwavelength
separation was used to modulate the phase of incident light along the interface.
It was demonstrated that, depending on the designed phase gradient, the
refraction angle can be controlled at will for any incident angle.
The authors of~\cite{2002_Enoch} showed that a metamaterial with a near-to-zero
refractive index acts as an antenna with an extremely high directivity --- a
source embedded in a slab of such metamaterial emits the waves whose refraction
at the interfaces with the surrounding media causes concentration of the outgoing
energy in a narrow cone.
A phenomenon that is reverse to this directive emission was predicted by Feng
in~\cite{2012_Feng}.
Feng showed that the direction of the incoming energy flow bends towards
the interface normal for any incident angle when p-polarized (transverse
magnetic) light enters an $\e$-near-zero metamaterial (the metamaterial with
vanishingly small real part of permittivity).
He clarified that such all-angle collimation of the incident light (in the
original paper, term "omnidirectional bending" was used) is a result of material
losses.

In our work we show that similar all-angle collimation can be realized with
s-polarized (transverse electric) incident light at the interface of
metamaterials with a vanishingly small real part of permeability, so-called
$\m$-near-zero metamaterials.
Note, however, that the theoretical framework we use is different from the one
employed by Feng~\cite{2012_Feng}:
To obtain our result we apply the theory of inhomogeneous waves which is commonly
used to describe refraction of light in lossy
media~\cite{2004_Fedorov,1983_Chen}.
Such approach allows us to generalize the result by Feng~\cite{2012_Feng} and
shed some light on its polarization dependence.
We show that all-angle collimation of incident light in $\e$-near-zero and
$\m$-near-zero metamaterials is the manifestation of the same phenomenon which
takes place under different polarization conditions.

A vanishingly small real part of permeability can be found in negative-index
metamaterials near permeability resonances which are used to achieve a negative
index of refraction.
To confirm our idea about all-angle collimation of incident light in
$\m$-near-zero metamaterials, we calculate the transmission angle of the Poynting
vector at the interface with the negative-index metamaterial recently reported by
Garc{\'\i}a-Meca {\it et~al.}~\cite{2011_Garcia-Meca}.

%********************************************************************************
\section{Inhomogeneous waves in a lossy metamaterial} \label{sec:2}
%********************************************************************************

Since all-angle collimation of incident light is a consequence of losses in a
medium~\cite{2012_Feng} and the propagation of light in lossy media differs 
from that in lossless media, we first summarize the basic features of light waves 
in lossy media~\cite{2004_Fedorov,1983_Chen,Fedorov}. Unlike the case of light 
waves in lossless media, the equiamplitude and equiphase planes of light waves 
in lossy media are not parallel, and such waves are called {\it inhomogeneous
waves}. 
The summarized results in this section will be used in the following sections 
to determine the direction of the Poynting vector of the wave transmitted 
through a metamaterial.

\begin{figure}[t]
    \centering
    \includegraphics{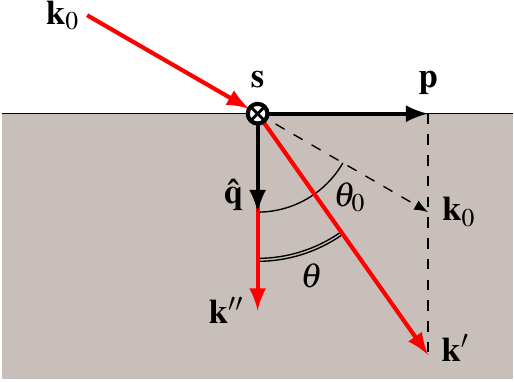}
    \caption{\label{fig:fig1}%
             Refraction of the wave at the interface of a lossy material.
             The incident plane wave has a real wave vector $\vc{k}_0$, while
             the wave vector $\vc{k}$ of the transmitted wave is complex due to
             material losses: $\vc{k}=\vc{k}'-i\vc{k}''$.
             In general, the phase $\vc{k}'$ and attenuation $\vc{k}''$ vectors
             are not parallel, and the transmitted wave is inhomogeneous.} 
\end{figure}

We consider an interface between two isotropic media (see Fig.~\ref{fig:fig1}).
The first medium is a lossless dielectric with a real refractive index $n_0$ and
the second medium is a lossy metamaterial with complex permittivity
$\e=\e'-i\e''$ and permeability $\m=\m'-i\m''$.
An incident plane wave with a real wave vector $\vc{k}_0$ comes from the first
medium.
The incident angle $\theta_0$ is an angle between $\vc{k}_0$ and the unit vector
normal to the interface $\uv{q}$, which is pointing to the second medium.
The complex electric and magnetic fields, $\vc{E}$ and $\vc{H}$, respectively, of
the transmitted wave are written as
\begin{equation} \label{eq:EH}
    \vc{E} = \vc{e}\exp[i(\w t-\vc{k}\cdot\vc{r})], \qquad
    \vc{H} = \vc{h}\exp[i(\w t-\vc{k}\cdot\vc{r})],
\end{equation}
where $\vc{e}$ and $\vc{h}$ are complex amplitude vectors, and $\vc{k}$ and 
$\vc{r}$ are a wave vector and a position vector, respectively, with $\w$ and 
$t$ being the wave frequency and time.
Since the second medium is lossy, the wave vector of the transmitted wave is
complex: $\vc{k}=\vc{k}'-i\vc{k}''$, where $\vc{k}'$ and $\vc{k}''$ are the real
phase and attenuation vectors, respectively, and they are written as 
\begin{equation} \label{eq:k}
    \vc{k}' = \vc{p} + q'\uv{q}, \qquad
    \vc{k}'' = q''\uv{q}.
\end{equation}
In Eq.~\eqref{eq:k} the phase vector $\vc{k}'$ is decomposed as a sum of two
vectors, namely, $\vc{p}$ and $q'\uv{q}$ (see Fig.~\ref{fig:fig1}).
Vectors $\vc{p}=[\uv{q}\times[\vc{k}'\times\uv{q}]]$ and $q'\uv{q}$ are,
respectively, parallel and perpendicular to the interface.
The attenuation vector $\vc{k}''$ is always normal to the interface.
Thus vectors $\vc{k}'$ and $\vc{k}''$ are not parallel (the only exception is the
case of normal incidence, $\vc{p}=0$) and the transmitted wave is inhomogeneous
as mentioned earlier.
The normal components of vectors $\vc{k}'$ and $\vc{k}''$ have magnitudes
$q'=(\vc{k}'\cdot\uv{q})$ and $q''=(\vc{k}''\cdot\uv{q})$, respectively, which
are given by~\cite{Fedorov}
\begin{equation} \label{eq:q}
    q'  = \sgn\{\xi''\}\frac{\w}{c_0}\sqrt{(|\xi|+\xi')/2}, \qquad
    q'' =\frac{\w}{c_0}\sqrt{(|\xi|-\xi')/2},
\end{equation}
where $\xi'=\e'\m'-\e''\m''-n_0^2\sin^2\theta_0$, $\xi''=\e'\m''+\e''\m'$,
$|\xi|=\sqrt{\xi'^2+\xi''^2}$, and $\sgn\{\xi''\}$ represents the sign of
$\xi''$.
Equation~\eqref{eq:q} says that the normal component $q'$ of the phase vector
$\vc{k}'$ is positive (i.~e., $\vc{k}'$ is directed away from the interface) if
$\xi''>0$, and negative (i.~e., $\vc{k}'$ is directed towards the interface) if
$\xi''<0$.
The latter case corresponds to negative refraction~\cite{Fedorov}.

The refractive index $m'$ and the attenuation coefficient $m''$ are formally
defined by $|\vc{k}'|=m'\frac{\w}{c_0}$ and $|\vc{k}'|=m''\frac{\w}{c_0}$.
In practice they can be found from the following equations:
\begin{equation} \label{eq:m}
    m'  = \sqrt{(|\xi|+\xi'+2n_0^2\sin^2\theta_0)/2}, \qquad
    m'' = \sqrt{(|\xi|-\xi')/2}.
\end{equation}
Note that, according to Eq.~\eqref{eq:m}, both $m'$ and $m''$ depend on the
incident angle $\theta_0$.
The transmission angle $\theta$ is the angle between $\vc{k}'$ and $\uv{q}$, and
can be found from the Snell's law, $m'\sin\theta=n_0\sin\theta_0$.

%********************************************************************************
\section{Poynting vector}
%********************************************************************************

In this section we derive the equation for the Poynting vector of the wave
transmitted into a metamaterial.
For this purpose we first consider the decomposition of the complex amplitude
vectors $\vc{e}$ and $\vc{h}$ of the transmitted wave into the s- and p-polarized
modes (TE and TM modes, respectively).
Since the plane of incidence is a plane spanned by vectors $\vc{k}_0$ and
$\uv{q}$, the vector $\vc{s}=[\vc{k}_0\times\uv{q}]=[\vc{k}\times\uv{q}]$ is
normal to the plane of incidence (see Fig.~\ref{fig:fig1}).
Using vector $\vc{s}$, we can decompose the complex electric vector amplitude 
$\vc{e}$ into the s- and p-polarized components: $\vc{e}=\vc{e}_s+\vc{e}_p$, 
where 
$\vc{e}_s=\vc{s}^{-2}(\vc{e}\cdot\vc{s})\vc{s}$ and
$\vc{e}_p=\vc{s}^{-2}[\vc{s}\times[\vc{e}\times\vc{s}]]=
 \vc{s}^{-2}(\vc{e}\cdot\uv{q})[\vc{s}\times\vc{k}]$, or
\begin{equation} \label{eq:e}
    \vc{e} = A_s\vc{s}+A_p[\vc{s}\times\vc{k}],
\end{equation}
where $A_s=\vc{s}^{-2}(\vc{e}\cdot\vc{s})$ and
$A_p=\vc{s}^{-2}(\vc{e}\cdot\uv{q})$ are the complex amplitudes of the s- and
p-polarized components, respectively.
Amplitudes $A_s$ and $A_p$ of the transmitted wave are connected with the
corresponding amplitudes of the incident wave by Fresnel coefficients
\cite{2004_Fedorov,1983_Chen}, i.~e., $A_p=0$ if the incident wave is
s-polarized, and $A_s=0$ if it is p-polarized.
To find the decomposition of the complex magnetic vector amplitude $\vc{h}$, we
substitute Eq.~\eqref{eq:e} into the identity
$\vc{h}=(\m_0\m\w)^{-1}[\vc{k}\times\vc{e}]$, and obtain
\begin{equation} \label{eq:h}
    \vc{h} = \e_0\e\w A_p\vc{s} - \frac{A_s}{\m_0\m\w}[\vc{s}\times\vc{k}].
\end{equation}

Now, with the help of Eq.~\eqref{eq:EH}, we write a complex time-averaged
Poynting vector as $\vc{S}=\frac{1}{2}[\vc{E}\times\vc{H}^*]=
\frac{1}{2}[\vc{e}\times\vc{h}^*]\exp[-2(\vc{k}''\cdot\vc{r})]$, where "$^*$"
means complex conjugate.
Substituting Eqs.~\eqref{eq:e} and \eqref{eq:h} into the last expression, we find
that vector $\vc{S}$ can be written as a sum of three components,
$\vc{S}=\frac{1}{2}(\vc{S}_s+\vc{S}_p+\vc{S}_{sp})\exp[-2(\vc{k}''\cdot\vc{r})]$,
where
\begin{subequations} \label{eq:S}
\begin{align}
    \vc{S}_s & = \frac{|A_s|^2}{\m_0\m^*\w}
                 [\vc{s}\times[\vc{k}^*\times\vc{s}]], \\
    \vc{S}_p & = \e_0\e^*\w|A_p|^2[\vc{s}\times[\vc{k}\times\vc{s}]], \\
    \vc{S}_{sp} & = -\frac{A_s^*A_p}{\m_0\m^*\w}
                    [[\vc{k}\times\vc{s}]\times[\vc{k}^*\times\vc{s}]].
                    \label{eq:Ssp}
\end{align}
\end{subequations}
The s-polarized component $\vc{S}_s$ depends only on $A_s$, while the p-polarized
component $\vc{S}_p$ depends only on $A_p$.
The cross-polarized component $\vc{S}_{sp}$ depends on both $A_s$ and $A_p$.
According to Eq.~\eqref{eq:Ssp}, the cross-polarized component $\vc{S}_{sp}$
exists only in lossy media where the wave vector $\vc{k}$ is complex.

A real time-averaged Poynting vector $\vc{P}$ corresponds to the real part of
$\vc{S}$.
Expanding the vector products and taking the real parts of Eq.~\eqref{eq:S}, we
find that
$\vc{P}=\frac{1}{2}(\vc{P}_s+\vc{P}_p+\vc{P}_{sp})\exp[-2(\vc{k}''\cdot\vc{r})]$,
where
\begin{subequations} \label{eq:P}
\begin{align}
    \vc{P}_s & = \frac{\vc{s}^2|A_s|^2}{\m_0|\m|^2\w}
                 (\m'\vc{k}'+\m''\vc{k}''), \\
    \vc{P}_p & = \e_0\w\vc{s}^2|A_p|^2(\e'\vc{k}'+\e''\vc{k}''), \\
    \vc{P}_{sp} & = \frac{2q''\vc{s}^2}{\m_0|\m|^2\w}
                    (\m'\Im\{A_s^*A_p\}-\m''\Re\{A_s^*A_p\})\vc{s}
\end{align}
\end{subequations}
with $\Re\{\}$ and $\Im\{\}$ being the real and imaginary parts of the
corresponding expressions.
Equation~$\eqref{eq:P}$ says that the s- and p-polarized components of the
Poynting vector are proportional to the sum of vectors $\vc{k}'$ and $\vc{k}''$,
while the cross-polarized component $\vc{P}_{sp}$ is proportional to vector
$\vc{s}$ and thus normal to the plane of incidence.
The component $\vc{P}_{sp}$ is responsible for the transversal shift of the
transmitted light beam.
A similar shift named Imbert-Fedorov shift takes place for the reflected light
beam for the case of total internal reflection \cite{1972_Imbert,2013_Fedorov}.
Despite the fact that these two shifts look similar, there are several different
viewpoints on the component $\vc{P}_{sp}$:
The authors of~\cite{1981_Halevi} argue that "there is no mechanism for energy
transport in the direction perpendicular to the plane of incidence" and set
$\vc{P}_{sp}$ equal to zero, based on the fact that "the Poynting vector is
defined only up to an arbitrary, additive, solenoidal vector".
Fedorov in his book \cite{2004_Fedorov} believes that this component is real and
responsible for the light pressure in the direction perpendicular to the plane of
incidence.
The authors of~\cite{2011_Dmitruk} came to the conclusion that the appearance of
$\vc{P}_{sp}$ is caused by excitation of surface electric polariton mode or
surface magnetic mode by the resonant or non-resonant manner.
In this work, however, we only consider s- ($A_p=0$) or p-polarized ($A_s=0$)
incident wave for which $\vc{P}_{sp}=0$.

By inspecting Eq.~\eqref{eq:P}, we find that both $\vc{P}_s$ and $\vc{P}_p$ are
parallel to $\vc{k}''$ if $\m'$ or $\e'$ is equal to zero, respectively.
Meanwhile, for any incident angle, the attenuation vector $\vc{k}''$ is always
normal to the interface [see Eq.~\eqref{eq:k}].
Therefore, we conclude that, at the interface of a material with a vanishingly
small real part of permeability ($\m'=0$), the s-polarized incident wave gives
rise to the transmitted wave whose energy flow is directed normally to the
interface, irrespective of the incident angle.
A similar argument holds for the p-polarized incident wave at the interface of
a material with a vanishingly small real part of permittivity ($\e'=0$), as shown
by Feng~\cite{2012_Feng}.

In conclusion we would like to note that it is possible to express the Poynting
vector in Eq.~\eqref{eq:P} in terms of the s- and p-polarized components of the
magnetic field $\vc{H}$.
The easiest way to do so is to apply the usual electromagnetic duality by
replacing $\e_0$ and $\e$ by $\m_0$ and $\m$, and using the corresponding
amplitudes $B_s$ and $B_p$ ($\vc{h} = B_s\vc{s}+B_p[\vc{s}\times\vc{k}]$) instead
of $A_s$ and $A_p$.
After such procedures we will find that all-angle collimation in $\e$-near-zero
and $\m$-near-zero metamaterials happens for s- and p-polarized magnetic field
$\vc{H}$, respectively, which is opposite to the case of $\vc{E}$.
However, this is not surprising, since vectors $\vc{E}$ and $\vc{H}$ are
perpendicular in the incoming wave, and, for example, s-polarization in terms of
$\vc{E}$ corresponds to p-polarization in terms of $\vc{H}$.

%********************************************************************************
\section{Transmission angles}
%********************************************************************************

\begin{figure}[t]
    \centering
    \includegraphics[width=7cm]{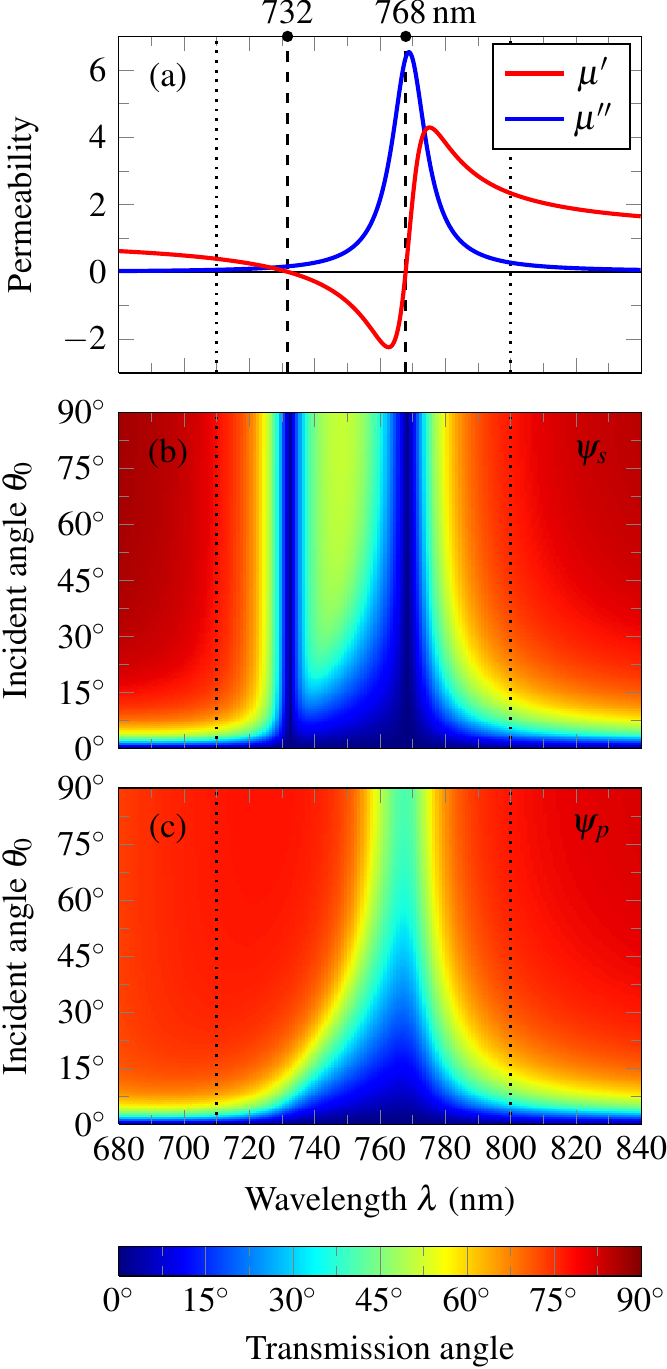}
    \caption{\label{fig:fig2}%
             (a) Variation of the real $\m'$ and imaginary $\m''$ parts of the
             permeability $\m$ as a function of wavelength $\lambda$.
             (b) and (c) The transmission angles $\psi_s$ and $\psi_p$ for the s-
             and p-polarized components of the Poynting vector as functions of
             wavelength $\lambda$ and incident angle $\theta_0$.
             All above functions are calculated for the interface between vacuum
             and the metamaterial reported in~\cite{2011_Garcia-Meca}.
             The spectral region of negative refraction is located between the
             two dotted vertical lines.} 
\end{figure}

Now we consider the transmission angles $\psi_s$ and $\psi_p$ for s- and
p-polarized components of the Poynting vector. 
Angles $\psi_s$ and $\psi_p$ are defined as the angles between vectors $\vc{P}_s$
and $\vc{P}_p$, respectively, and the unit normal $\uv{q}$.
Consider, for example, the angle $\psi_s$.
We can find this angle from the equation,
$\tan\psi_s=|[\vc{P}_s\times\uv{q}]|/(\vc{P}_s\cdot\uv{q})$.
Using Eq.~\eqref{eq:P} and taking into account that
$[\vc{k'}\times\uv{q}]=\vc{s}$ and $\vc{k}''\parallel\uv{q}$, we find that
$\tan\psi_s=\m'|\vc{s}|/(\m'q'+\m''q'')$.
Similarly we obtain $\tan\psi_p=\e'|\vc{s}|/(\e'q'+\e''q'')$.
This form of equations for $\psi_s$ and $\psi_p$ was previously obtained
in~\cite{1981_Halevi}.
Using the equations of $|\vc{s}|=m'\frac{\w}{c_0}\sin\theta$,
$q'=m'\frac{\w}{c_0}\cos\theta$, and $q''=m''\frac{\w}{c_0}$, we finally obtain
\begin{equation} \label{eq:tsp}
    \tan\psi_s = \frac{\m'm'\sin\theta}{\m'm'\cos\theta+\m''m''}, \qquad
    \tan\psi_p = \frac{\e'm'\sin\theta}{\e'm'\cos\theta+\e''m''}.
\end{equation}
Here we see another manifestation of all-angle collimation.
Namely, in case of $\m'=0$ or $\e'=0$ the corresponding transmission angle,
$\psi_s$ or $\psi_p$, is equal to zero, irrespective of the incident angle.
Moreover, Eq. \eqref{eq:tsp} says that the transmission angles $\psi_s$ and
$\psi_p$ are not equal, which means that the direction of the energy flow in a
lossy material is different for s- and p-polarized incident
wave~\cite{1981_Halevi}.
However, for any natural material this difference is negligibly small, since
usually $\m''/\m'\ll1$ and $\e''/\e'\ll1$, which, according to
Eq.~\eqref{eq:tsp}, means that $\psi_s\simeq\psi_p\simeq\theta$.
Nevertheless, the difference between $\psi_s$ and $\psi_p$ can be significant in
metamaterials where the above inequalities may not hold.

To be more quantitative we calculate the values of $\psi_s$ and $\psi_p$ for the
case in which the first medium is vacuum ($n_0=1$) and the second medium is the
negative-index metamaterial reported in~\cite{2011_Garcia-Meca}.
We retrieve the relevant parameters for the permittivity $\e$ and permeability
$\m$ of this metamaterial, performing the parameter fitting for the Drude
model~\cite{2011_Fedorov}.
Figure \ref{fig:fig2}(a) shows the dependence of the real and imaginary parts of
the retrieved permeability $\m$ on the wavelength $\lambda$.
We see that this dependence has a resonance feature and the real part $\m'$ of
the permeability is equal to zero at the wavelengths 732 and 768\,nm.
At these wavelengths we expect all-angle collimation for incident s-polarized
wave.

Using the retrieved functions for $\e$ and $\m$, we calculate $\psi_s$ and
$\psi_p$ by Eq.~\eqref{eq:tsp} where values of $m'$, $m''$, and $\theta$ have
been obtained using the equations in section \ref{sec:2}.
Figures \ref{fig:fig2}(b) and \ref{fig:fig2}(c) show the dependencies of the
transmission angles $\psi_s$ and $\psi_p$ on the wavelength $\lambda$ and the
incident angle $\theta_0$.
In spite of negative refraction, we consider $\psi_s$ and $\psi_p$ as angles
between two vectors and set them positive.

As expected, we see in Fig.~\ref{fig:fig2}(b) that, at wavelengths where $\m'=0$,
the transmission angle $\psi_s$ is zero for any incident angle.
Therefore, the direction of energy flow in the second medium will be normal to
the interface for any incident angle.

By comparing Figs.~\ref{fig:fig2}(b) and \ref{fig:fig2}(c), we clearly see the
difference between $\psi_s$ and $\psi_p$.
This difference is more significant at the wavelength $\lambda=732$\,nm where we
have all-angle collimation for the s-polarized component of the Poynting vector.
We hope that this observation will encourage experimentalists to verify the
difference between the transmission angles for s- and p-polarized incident light.

Before proceeding to the conclusions, we would like to briefly discuss the case
in which light propagates to the reverse direction, that is, from $\e$-near-zero
or $\m$-near-zero lossy metamaterial into an ordinary medium.
One may naively think that we will have a directive emission similar to the one
in~\cite{2002_Enoch}.
However, a simple generalization of our result to the reverse problem leads to an
unphysical solution:
In this work we have assumed that a homogeneous (the attenuation vector equals
zero) plane wave comes from a lossless medium to the interface of a lossy
metamaterial.
Therefore, the most natural way to formulate the reverse problem is to assume
that a homogeneous damped (the nonzero attenuation vector is parallel to the wave
vector) plane wave comes from a lossy metamaterial to the interface with a
lossless medium.
This assumption implies that somewhere deep in the metamaterial there is an
embedded light source, whose radiation, close to the interface, can be
approximated by homogeneous damped waves.
Upon arrival to the interface, the attenuation vector of these homogeneous damped
waves will have a nonzero tangential component (the only exception is the waves
propagating along the normal to the interface).
Since the tangential components of attenuation vectors are continuous across any
interface~\cite{Fedorov}, the wave outgoing from the metamaterial will have
nonzero attenuation vectors, that is, the wave transmitted into the lossless
surrounding medium will be inhomogeneous.
A closer examination shows that the attenuation vector of the outgoing wave will
point towards the interface, which means that the amplitude of the outgoing wave
will grow unlimitedly, which is unphysical.
Therefore, we conclude that the radiation of a light source embedded in a lossy
metamaterial can not be represented by homogeneous damped waves; to find a
correct solution one should consider the mechanism of light emission from such
source in more details, which is out of scope of this work.

%********************************************************************************
\section{Conclusions}
%********************************************************************************

In conclusion we have theoretically studied the transmission of light waves in
$\m$-near-zero metamaterials.
Similar to the case of $\e$-near-zero metamaterials~\cite{2012_Feng}, we have
found the effect of all-angle collimation of incident light in $\m$-near-zero
metamaterials for the s-polarized incident wave.
Thus we have provided the generalized footing, based on which we can show, with
sufficient clarity, that all-angle collimation of incident light in
$\e$-near-zero and $\m$-near-zero metamaterials is the manifestation of the same
phenomenon under different polarization of incident light.
We have presented specific results for the negative-index metamaterial with a
permeability resonance where the real part of the permeability becomes zero.
Additionally, we have shown that the transmission angle of the Poynting vector
depends on the polarization of the incident wave, and this difference is very
significant in the spectral region where all-angle collimation of incident light
takes place.

%********************************************************************************
\section*{Acknowledgments}
%********************************************************************************

This work was supported by a Grant-in-Aid for Scientific Research from the
Ministry of Education and Science of Japan.
Part of the work by V.\,Yu.~Fedorov was also supported by the Japan Society for
the Promotion of Science (JSPS).

\end{document}